\documentclass{article}
\usepackage[utf8]{inputenc}
\usepackage[T1]{fontenc}
\usepackage{graphicx}
\usepackage{grffile}
\usepackage{longtable}
\usepackage{wrapfig}
\usepackage{rotating}
\usepackage[normalem]{ulem}
\usepackage{amsmath}
\usepackage{textcomp}
\usepackage{amssymb}
\usepackage{hyperref}
\usepackage{listings}
\author{Phil Scott, Steven Obua, Jacques Fleuriot}
\date{}
\title{Bootstrapping LCF Declarative Proofs}
\hypersetup{
 pdfauthor={Phil Scott, Steven Obua, Jacques Fleuriot},
 pdftitle={Bootstrapping LCF Declarative Proofs},
 pdfkeywords={},
 pdfsubject={},
 pdfcreator={Emacs 25.1.1 (Org mode 9.0.5)}, 
 pdflang={English}}
\begin{document}

\maketitle
\begin{abstract}
Suppose we have been sold on the idea that formalised proofs in an LCF system should
resemble their written counterparts, and so consist of formulas that only provide
signposts for a fully verified proof. To be practical, most of the fully elaborated
verification must then be done by way of general purpose proof procedures. Now if
these are the only procedures we implement outside the kernel of logical rules, what
does the theorem prover look like? We give our account, working from scratch in the
\emph{ProofPeer} theorem prover~\cite{ProofPeer}, making observations about this new setting along the way.
\end{abstract}

\section{Introduction}
\label{sec:orgdf06c06}

LCF~\cite{LCF} style theorem provers are based on small kernels of inference/typing
rules. Ultimately, all theorems are proven by applying these rules, though in
practice, users will call derived inference rules and higher-level proof procedures
which call yet more derived inference rules and procedures until they ultimately hit
the kernel rules. This leads to a zoo of often composable proof tools
culminating in full decision procedures, which together afford the user a variety of
ways to verify their theorems balancing elegance and simplicity against performance.

There is a more consolidated approach to verification with a less programmatic
aesthetic. One can use a \emph{declarative language}, such as those offered by Isabelle,
HOL~Light and Coq \cite{DeclarativeCoq,Isar,MizarLight}. A user of such a
language writes structured proofs by stating assumptions, variable declarations and
intermediate conclusions, and indicating how these are connected by logical
consequence. The languages often approximate written mathematical proofs, and could
be viewed as the ideal implementation of a formal language for written mathematics.
We might then expect them to be more accessible to working mathematicians, a core
demographic for ProofPeer. Thus, the language with which users interface with our
system, \emph{ProofScript}~\cite{ProofScript}, takes structured proof as a core part of its syntax, with the
traditional panoply of proof procedures intended only to support these readable
declarative proofs.

In both Isar and HOL~Light's Mizar~Light, the heavy duty automation tends to
be handed to clausal form automated theorem provers, and we have not
sought to make ProofPeer an exception. In the following sections, we describe our
streamlined approach to implementing just the scaffolding necessary to make such
provers work in the novel setting of ProofScript, a dynamically typed, purely
functional language with primitives for working with the terms and theorems of the
safely encapsulated and secure logical kernel of ProofPeer.

\section{Organisation}
\label{sec:orgb15eba7}
In the next section, we will give some motivation for the use of general purpose,
external automated theorem provers in interactive theorem proving. In
\S\ref{sec:org6f28b5a}, we shall describe the minimal infrastructure
we used to talk the basic language of these provers and in
\S\ref{sec:orgb5ce798} we shall discuss what we needed to
understand the response. We conclude in \S\ref{sec:orgf5b06aa}.

\section{Accessing external tools}
\label{sec:orgba69b1f}

The Sledgehammer~\cite{IsabelleSledgehammer} tool in Isabelle began as an impressive effort to integrate external
theorem provers, initially Vampire~\cite{Vampire} and SPASS~\cite{SPASS}, into declarative proofs by
automatically processing their certificates as inference rules in the Isar
language. Since then, many more theorem provers have been integrated~\cite{Sledgehammer2015}, but it has
turned out that it is sufficient, and much easier besides, to just use the provers to
minimise the number of lemmas needed for a first-order proof and then let it be found
by Hurd's \texttt{METIS}~\cite{metis}. \texttt{METIS} is a generic resolution prover for
first-order logic with equality, based on a small LCF style kernel with full proof-recording. As with other resolution
provers, its job is to refute the negation of a posed problem, and because of its
easily trusted kernel, when \texttt{METIS} claims to find a refutation, we have a high
guarantee that the proof record really is a proof of the refutation.

Both Isabelle and \texttt{METIS} are written in Standard ML, so the automated theorem prover
is easily usable as a library of the other interactive theorem prover. ProofPeer, on
the other hand, runs on the JVM and Javascript, and is implemented entirely in
Scala. One of our first efforts, then, was to write a clean port of \texttt{METIS} to Scala,
where it can now be used as a library for other JVM or Javascript
based-projects. Entirely under our control, this prover is in place to become the
standard automated theorem prover which ProofPeer uses to reconstitute first-order
proofs, just as its Standard ML variant does for Isabelle's Sledgehammer.

To access \texttt{METIS} from ProofScript, we equip the language with a built-in function
\texttt{callmetis} which can send and receive ProofScript data structures to and from a
wrapper around the \texttt{METIS} library. This is currently a somewhat \emph{ad-hoc} approach, but
ProofScript's data structures are easy to read and write and can serve as a general data
exchange format in the manner of S-expressions and JSON, and we certainly intend to
provide a general mechanism to integrate more external tools with ProofScript,
communicating with wrappers in the same way.

A cautionary note is due because functions which get their data from third parties
may not behave nicely and return the same output for the same input, thereby breaking
ProofScript's promises about functional purity. For now, we can at least be reassured
that our implementation of \texttt{METIS} is guaranteed to be fully deterministic, and that
\texttt{callmetis} will always return the same output for the same input.

\section{Conversion}
\label{sec:org6f28b5a}
Any time a user wishes to make an inference in a declarative proof, they will
implicitly have in mind a conjecture that the next formula to be derived is a logical
consequence of the assumptions currently in force. To a resolution prover, this is
turned into a conjunctive normal form problem asserting that the contradiction of the
next step together with the assumptions in force entails a contradiction. It
therefore suffices to show that such a refutation entails the user's conjecture.

Normally, however, we show something much stronger, namely that the user's conjecture
and the refutation of its conjunctive normal form are logically equivalent, by
computing the latter from the former by successive term rewriting. From its earliest
days, the original LCF theorem prover implemented its rewriters via conversions~\cite{LCFRewriting},
which still represent an impressive demonstration of how higher-order functions can
be used to design combinator languages~\cite{CombinatorLanguages}. ProofScript has excellent support for
higher-order functions, and even though it eschews the static type system familiar to
users of Standard ML, we found this gave us no real difficulty when it came to
implementing and using conversions.

In the next few sections, we shall discuss how we go about building the basic
infrastructure to compute the CNF form of the user's conjecture. We shall then
describe how we send the version of the problem to \texttt{METIS} via \texttt{callmetis} and how we
interpret \texttt{METIS}'s resolution proof trace from its own kernel.

\subsection{Primitive inference rules}
\label{sec:org8fd0c9c}
Our kernel implements rules for a classical, monomorphic, simply typed lambda
calculus with just two primitive types: the type of propositions \(\mathbb{P}\) and the
type \(\mathcal{U}\) of ZFC sets. All theorems generated by this kernel are associated
with rich context objects, which provide the hypotheses of the theorem, and
information on how to resolve names to fully qualified identifiers with a logical
type.

The contexts are used to support structured declarative proof. At any point in
ProofScript code, there is an assumed context of constants that have been introduced
and assumptions which have been made of these constants, which can be extended via
primitive \texttt{assume} and \texttt{let} statements. ProofScript code is then structured in
nested blocks, allowing us to escape back into a parent context as we leave each code
block.

The use of these contexts means that there is no need to distinguish constants from
free variables, as we do in traditional HOL based theorem provers. In ProofScript, we
do not instantiate free variables in terms, but only specialise quantifiers. We
believe this makes ProofScript conceptually simpler for end-users, though as we shall
discuss in the coming sections, it presents some complications as far as interfacing
with our resolution prover, which does make such syntactic distinctions.

\subsection{Conversions and conversionals}
\label{sec:org90e0807}
Historically, conversions were functions which take a term \(t\) and either produce an
equational theorem \(t = t'\) or else throw an exception. These exceptions are
standardly caught by derived conversions which may try an alternative conversion
strategy. We do not support such an exception mechanism in ProofScript, so our
conversions must signal any errors via appropriate return values.

Conversions are operated upon using higher-order functions called \emph{conversionals},
which allow us to build conversion strategies out of existing conversions, targeting
individual parts of a term and responding to failure by trying alternative
strategies.

Most of the 1500 lines of ProofScript library code which integrates ProofScript with
\texttt{METIS} does so by way of conversions. We define about 50 conversions and conversionals
which are used around 500 times.

\subsubsection{Basic conversions}
\label{sec:org7402f3a}
Some basic conversions are mostly implemented directly in terms of the kernel's
inference rules or via simple derived rules. Other conversionals exist simply to
deal with failure. The conversion \(c_1\) \texttt{orElseConv} \(c_2\) takes a term \(t\) and
tries \(c_1\;t\), checks if it fails, and if so, returns \(c_2\;t\) instead. Another
conversion, \texttt{noConv}, is the identity of \texttt{orElseConv}, and is just the constant
function to \texttt{nil}, representing a conversion that always fails.

The conversional \texttt{thenConv} provides sequencing, taking two conversions
\mbox{$t\rightsquigarrow t'$} and \(t' \rightsquigarrow t''\), and producing the
conversion \(t \rightsquigarrow t''\) via a derived rule for transitivity of
equality. This conversional also has an identity, \texttt{allConv}, which is just an alias
for the kernel's reflexivity inference rule.

As noticed by Paulson~\cite{LCFRewriting}, both conversionals are associative, with \texttt{thenConv}
distributing over \texttt{orElseConv} on the left. We also notice that, if we make
conversions non-deterministic such that they return lists of equations with the
empty list representing failure, we retain right-distributivity, while if we
have them return sets, we get all the axioms of a ring.

\subsubsection{Subterm conversionals}
\label{sec:orgc277e44}
ProofScript terms compose only as combinations and abstractions, and so it
suffices to define two basic conversionals which can be used to arbitrarily
rewrite subterms. The conversional \texttt{combConv} is implemented in terms of the
primitive kernel destructor \texttt{destcomb}, which splits combinations, and the
inference rule \texttt{combine} which says that combining equal terms yields equal
combinations. The conversional accepts conversions \(f \rightsquigarrow g\) and \(x
\rightsquigarrow y\) and yields the conversion \(f\;x \rightsquigarrow g\;y\). 

Another conversional deals with abstractions via the primitive kernel destructor
\texttt{destabs}, which tears apart abstractions, and the inference rule \texttt{abstract} which
sends \(\forall x.\; \phi(x) = \psi(x)\) to \mbox{$(x \mapsto \phi(x)) = (y \mapsto\psi(y))$}.  
Using these, \texttt{absConv} sends the conversion \(\phi(x) \rightsquigarrow \psi(x)\) to the
conversion \mbox{$x \mapsto \phi(x) \rightsquigarrow x \mapsto \psi(x)$}.

It should be noted that ProofScript's abstraction destructor differs from those of
traditional HOL systems, which would normally return the body of the abstraction as a
term with a new free variable. As mentioned in \S\ref{sec:org8fd0c9c},
all ProofScript variables are bound, leaving only the constants. Thus, when
passed an abstraction, \texttt{destabs} returns a triple \((v,\phi(v),\texttt{ctx})\),
where \texttt{ctx} extends the current context with a new declaration for a new constant
\(c\), and such that \(\phi(c)\) is a valid term in this context.

By locally entering the returned context \texttt{ctx}, we can apply our conversion \(\phi(x)
\rightsquigarrow \psi(x)\) which, if successful, yields the equational theorem
\(\phi(c) = \psi(c)\). When this is returned to the enclosing context, the kernel
discharges the constant declaration and replaces the constant in the theorem with a
universally quantified variable. We can then apply the \texttt{abstraction} inference rule
to yield our desired result.

\lstset{label= ,caption= ,captionpos=b,numbers=none}
\begin{lstlisting}
def absConv conv =
  tm =>
    match destabs tm
      case [ctx, _, body] =>
        match incontext <ctx> conv body
          case nil => nil
          case th => abstract (lift! th)
      case _ => nil
\end{lstlisting}

\subsubsection{Other primitive conversions}
\label{sec:org5a040c6}
As mentioned, ProofScript's kernel inference rule \texttt{reflexive} is already a
conversion. There is one other kernel rule which is a conversion, namely
\texttt{normalize}. This sends a term to its \(\beta\eta\) long normal form.

The following are the only other primitive conversions we need:

\begin{description}
\item[{\texttt{subsConv} }] Given a theorem \(\vdash t = t'\), sends \(t\) to a normalized \(t'\), and
any other term to \texttt{nil}.
\item[{\texttt{rewrConv1}}] Given a theorem 

$$\vdash \forall x_1\;x_2\;\ldots\;,x_n.\; \phi(x_1, x_2, \ldots,
                 x_n) = \psi(x_1, x_2, \ldots, x_n)$$

sends \(t\) to \(t'\) if it finds a substitution \(\theta\) such that \(t =
                 \phi(x_1, x_2, \ldots, x_n)[\theta]\) and \(t' = \psi(x_1, x_2,
                 \ldots, x_n)[\theta]\). If no such substitution is found, returns
\texttt{nil}~\footnote{The matching algorithm which finds \(\theta\) is quite
rudimentary, but sufficient for our purposes.}.
\end{description}

Of these, only the primitive rewriting conversion \texttt{rewrConv1} has a non-trivial
implementation in the form of a pattern matching algorithm for terms. The pattern is
represented by the left hand side of the quantified equation in the theorem passed to
\texttt{rewrConv1}. 

Here, there is some minor awkwardness in not having free variables. In a regular HOL,
 a pattern is straightforwardly represented by a term, with the substitutable atoms
 the free variables, and the unsubstitutable atoms the constants. ProofScript does
 not support free variables, so we must instead represent the substitutable parts of
 a pattern with bound variables and keep track of substitutions by associating
 new constants with binding positions.

The conversion \texttt{rewrConv1} therefore takes a quantified equational theorem whose
left-hand side is the pattern to match against. The quantified variables are the ones
we can substitute for, and the quantifiers are stripped by repeated applications of
\texttt{destabs}, giving us the equational body in a new context \texttt{ctx} where the quantified
variables have been replaced with fresh variables \(x_1, x_2, \ldots, x_n\) in the
quantification order. These variables will have been declared in \texttt{ctx}.

We then prime our matching algorithm with a list of those constants, so that it knows
it can treat these as substitutable while all other variables must be treated as
fixed constants. The matching algorithm returns an instantiation as an association
list \([(x_1, t_1), (x_2, t_2), \ldots, (x_n, t_n)]\). We can then specialise the
quantifiers of the original equational theorem in turn with \(t_1, t_2, \ldots, t_n\),
apply the equational form of Modus Ponens \footnote{This rule sends \(\vdash\phi = \psi\) and \(\vdash\phi\) to
\(\vdash\psi\).}, and thus obtain the desired right-hand side.

\subsubsection{Proving theorems by conversion}
\label{sec:org86d55ff}
By definition, a successful conversion proves a theorem given a term. But they can
also be used to infer non-equational theorems via \texttt{convRule}, which takes a theorem,
applies a conversion to its term, and if successful, uses the equational form of
Modus Ponens to prove the converted form of the theorem. We can also prove
non-equational theorems \(\vdash\phi\) is if we have a conversion \(\phi \rightsquigarrow
\top\). Equivalences of a proposition to true can be eliminated to the proposition via
a simple derived inference rule.

\subsection{CNF Conversion}
\label{sec:orgb9dc96d}

A full CNF conversion from ProofScript conjectures can be obtained through the
following passes:

\begin{itemize}
\item convert the term to first-order by:
\begin{itemize}
\item eliminating set builder notation;
\item lifting abstractions, by pulling them out to the front and making them the
arguments to \(\beta\) redexes (let-definitions);
\item lifting any terms of type \(\mathbb{P}\) used as first-order function or predicate
arguments, pulling them out to the front and making them the arguments to \(\beta\)
redexes;
\end{itemize}
\item converting the term to its negation normal form:
\begin{itemize}
\item eliminating all propositional connectives other than \(\wedge\), \(\vee\) and \(\neg\);
\item pushing all negations to the leaves of the term and eliminating double negations;
\end{itemize}
\item converting to prenex normal form by pulling all quantifiers out to the front of the
formula;
\item converting the propositional matrix of the quantified formula to conjunctive normal
form;
\item eliminating all quantifiers by skolemisation.
\end{itemize}

\subsection{Tautology checking}
\label{sec:org20a9f81}
The parts of the CNF conversion which manipulate propositional formulas are easy
enough to implement once we have proven a number of simple propositional
equivalences. These are tedious to obtain by hand, so it is helpful to be able to
decide and verify propositional tautologies automatically.

A simple but complete tautology checker can be implemented by what amounts to
truth-table evaluation. First, we collect a list of equational rewrite theorems, that
tell us the semantics of each propositional connective in terms of its valuations on
\(\top\) and \(\bot\) (theorems such as \(\vdash\forall p.\; (p \vee \top) = \top\)). Then,
given a propositional conjecture, we sweep through and extract each variable \(p\),
performing a case-analysis \(\vdash p = \top \vee p = \bot\). Each complete
case-analysis corresponds to a row of a truth-table, and gives us a set of equations
which evaluate each variable. Next, using a derived conversional \texttt{upConv}, we descend
to the leaves of a term and then work upwards, applying our equations and rewrite
rules, calculating the truth value of the whole proposition and hoping to reach the
\(\top\).

We thus have a tautology checker which can be used to prove classical propositional
theorems of a few variables (and at most three at a time suits our purposes). As a
decision procedure, it can be easily integrated into declarative proof by the
ProofScript keyword \texttt{by}, as in the following code:

\lstset{label= ,caption= ,captionpos=b,numbers=none,mathescape=true}
\begin{lstlisting}
theorem andDeMorgan: $\forall p\;q.\; (\neg(p \wedge q)) = (\neg p \vee \neg q)$
  by taut
\end{lstlisting}

Case splitting on propositional variables, as we have done, requires that we have
first proven the law of excluded middle, which can be done via Dionconescu's~\cite{AxiomChoiceExcludedMiddle} proof,
suitably modified for Hilbert's indefinite description operator, which assumes full
classical choice. ProofScript has a novel implementation of choice, and, unlike the
HOLs, does not take indefinite descriptions as primitive. The classical inference is
instead achieved automatically by ProofScript's rules for lifting theorems out of
contexts:

\lstset{label= ,caption= ,captionpos=b,numbers=none,mathescape=true}
\begin{lstlisting}
choose choiceDef: '$\epsilon\;:\;(\mathbb{P} \rightarrow \mathbb{P}) \rightarrow \mathbb{P}$'
  let '$p\;:\;\mathbb{P} \rightarrow \mathbb{P}$'
  assume ex:'$\exists x.\;p\ x$'
  choose 'chosen:$\mathbb{P}$' ex
\end{lstlisting}

Here, we define an \(\epsilon\) function based on a nested proof. This proof assumes
that there exists some object satisfying \(p\) and then \texttt{choose}\hspace{0cm}s a
witness. When the assertion that the witness satisfies \(p\) is lifted out of the
context, the assumptions are discharged and a quantifier introduced for the variable
\(p\), yielding the theorem
$$\forall p.\;\exists chosen.\; (\exists x.\;p\ x) \rightarrow p\ chosen.$$

By applying \texttt{choose} to this universal theorem, ProofScript will first skolemize to

$$\exists chosen.\; \forall p.\; (\exists x.\; p\ x) \rightarrow p\ (chosen\ p)$$
and then introduce \(\epsilon\) as a new Skolem constant.

\subsection{Skolemisation}
\label{sec:org03a3dd3}
A higher-order theorem which captures the main rewrite rule we need to perform
skolemisation is given by:

\begin{equation}\label{thm:skolem1}
  \vdash\forall p.\;(\forall x.\; \exists y.\;p\ x\ y) = (\exists f.\;\forall x.\; p\ x\
(f\ x)).
\end{equation}

Repeated rewriting with this theorem allows us to fully skolemize. To take an
example:

\begin{align*}
  \forall x\;y\;:\;\mathcal{U}.\;\exists z\;:\;\mathcal{U}.\; P\ x\ y\ z
  &= \forall x\;:\;\mathcal{U}.\;\exists f\;:\;\mathcal{U} \rightarrow \mathcal{U}.\;\forall y.\; P\ x\ y\ (f\ y)\\
  &= \exists g\;:\;\mathcal{U} \rightarrow \mathcal{U} \rightarrow \mathcal{U}.\;\forall x.\;\forall y.\; P\ x\ y\ (g\ y\ x).
\end{align*}.

But notice that the type of \(f\) in the second formula differs from the type of \(g\) in
the third, meaning that we are applying variants of \eqref{thm:skolem1}
with different types. This is no issue in traditional HOLs which support a simple
form of type polymorphism and type instantiation. ProofScript, however, due to a
deliberate decision to limit the expressivity of its higher-order metalogic, can only
express the above as a monomorphic theorem.

This means that an arbitrary number of variants of \eqref{thm:skolem1} may be
needed to skolemize theorems in ProofScript, each requiring its own proof, because
there are infinitely many possible types that the Skolem functions may inhabit. The
best we can hope to do is capture the theorems in a single definition, a metatheorem
expressed in the language of ProofScript, as it were:

\lstset{label= ,caption= ,captionpos=b,numbers=none,mathescape=true}
\begin{lstlisting}
def skolemThm [a,b] =
  theorem '$\forall p.\;(\forall x.\;\exists y.\; p\ x\ y) = (\exists f\;:\;\texttt{‹a›} \rightarrow \texttt{‹b›}.\;\forall x.\; p\ x\ (f\ x))$'
  $\ldots$
\end{lstlisting}

Here, ProofScript inserts the type variables passed as function arguments \texttt{a} and \texttt{b}
into the \texttt{‹ ›} quoted positions of the theorem.

This is still an unpleasant state of affairs, since every time we call \texttt{skolemThm},
we may find ourselves redoing a proof, namely when the arguments \texttt{a} and \texttt{b} have
been passed in before. Our solution is to introduce a primitive memoisation mechanism
via the ProofScript keyword \texttt{table}. This can replace \texttt{def} in the above code and will
cause ProofScript to save the result of calls to \texttt{skolemThm} and return the saved
copy whenever the function is called with the same arguments.

In general, it is worth being wary of the consequences of one's decisions to simplify
a logic because one does not have need of the expressive power. While we have chosen
to keep the logic monomorphic to encourage users to work as much as possible within
the ZF object logic, we would not want to push this to the point of using a
first-order logic, where we cannot properly express axioms such as the axiom of
comprehension, and where we would require the system to waste time proving
arbitrarily many instances of metatheorems for any generally useful consequence of
such axioms.

With conversions, the ProofScript code which performs the CNF conversion is
structurally very similar to the pseudocode algorithms which do the same, but with
the benefit that any result is formally verified for correctness. Conversions give us
a powerful way to drive \emph{computation}, and are well suited to this sort of
algebraic normalisation.

\section{Talking to \texttt{METIS}}
\label{sec:orgb5ce798}
\texttt{METIS}'s logical kernel deals only in CNF clauses, which are represented as sets of
literals. There are no connectives or quantifiers in the logical syntax.

\subsection{Representing \texttt{METIS} clauses and certificates}
\label{sec:org0719000}
\texttt{METIS} proof certificates are returned to ProofScript as a tree, rooted at the empty
clause which represents the refutation, and with nodes labelled with clauses derived
and the inference rule used. The possible rules are:

\begin{description}
\item[{\texttt{Axiom}}] leaf nodes containing an arbitrary clause;
\item[{\texttt{Assume}}] leaf nodes with clause of the form \(\{\phi, \neg\phi\}\);
\item[{\texttt{Refl}}] leaf nodes with clause of the form \(\{x = x\}\);
\item[{\texttt{Equality}\vspace{0cm}\((L(x),s,t)\)}] leaf nodes with a clause of the form \(\{ s \neq
     t, L', L[t/x]\}\) where \(L'\) is the negation of \(L[s/x]\);
\item[{\texttt{RemoveSym}}] a derived rule in Hurd's implementation, this node has a clause which
is the same as its child, minus duplicate equalities and
inequalities (up to symmetry of equality);
\item[{\texttt{Irreflexive}}] a node which is the same as its child, but without literals of the
form \(x \neq x\);
\item[{\texttt{Subst}\vspace{0cm}\((\theta)\)}] a node whose clause is a substitution \(\theta\) of
its child clause;
\item[{\texttt{Resolve}\vspace{0cm}\((L)\)}] uses of the resolution rule mark nodes with two
children, whose clauses are of the form \(C_1 \cup \{L\}\) and \(C_2 \cup \{L'\}\)
where \(L'\) is the negation of \(L\). The resolution's clause is \(C_1 \cup C_2 -
     \{L, L'\}\).
\end{description}

Generally, the more powerful \texttt{METIS}'s kernel rules, the more work we have cut out
for us on the ProofScript side. For one example, \texttt{METIS} implements clauses as sets
of literals, and so assumes a bunch of rules for working with disjunctions without
telling us how they are applied. For another, we have taken the liberty of replacing
some of \texttt{METIS}'s derived kernel rules with primitive ones, but each such liberty
comes at a cost of extra work in certificate translation.

To deal with literals as sets, it makes sense to ensure that our ProofScript clause
representation is in some normal form, and for us, that just means making sure we are
always normalised with respect to associativity and idempotency of disjunction. In
that way, we can treat our clauses as lists of literals with unique elements. Passes
over the clause to remove duplicates are only needed to implement \texttt{Subst} and
\texttt{Resolve}, and are handled with a conversion \texttt{nubClauseConv}. We need no pass for
associativity, which is always preserved by our implementation of the \texttt{METIS} kernel
rules.

In ProofScript, we represent proof certificates as a tree made by nesting lists, with
clauses as sets and literals encoded in an S-expression like format according to
\(\theta\):
\begin{displaymath}
  \theta(t) =
  \begin{cases}
    t, \quad\text{if $t$ is a variable}\\
    [F, \theta(t_1), \theta(t_2), \ldots, \theta(t_n)],
    \quad\text{if $t$ is the function application}\\
      \quad\quad\text{$F(t_1,t_2,\ldots,t_n)$ with function symbol $F$}\\
    [\texttt{true}, [P, \theta(t_1), \theta(t_2), \ldots, \theta(t_n)]],
    \quad\text{if $t$ is the positive literal}\\
      \quad\quad\text{$P(t_1,t_2,\ldots,t_n)$ with predicate symbol $P$}\\
    [\texttt{false}, [P, \theta(t_1), \theta(t_2), \ldots, \theta(t_n)]],
    \quad\text{if $t$ is the negative literal}\\
      \quad\quad\text{$\neg P(t_1,t_2,\ldots,t_n)$ with predicate symbol $P$}.
  \end{cases}
\end{displaymath}

Our port of \texttt{METIS} has literals which are polymorphic in variable, function symbol
and predicate symbol alphabets, and it is up to the caller to decide what alphabets
to use. So can get away with just using actual ProofScript function terms as \texttt{METIS}
function symbols and Proofscript predicate terms as predicate symbols.

\subsubsection{Free variables aren't}
\label{sec:org81faae6}
\texttt{METIS} variables often need to be instantiated, so when represented as a ProofScript
theorem, they must be bound in clauses by quantifiers. Since they must also be
instantiated on a clause-by-clause basis, we need a set of bindings per
clause. However, \texttt{METIS} likes to be able to identify variables across clauses, even
as it substitutes for them independently.

We cannot make this identification reliably in ProofScript, which only identifies
bound variables between clauses up to $\alpha$-equivalence, and even if we have some
control over the naming of variables, we cannot reliably control ProofScript's naming
of fresh ones.

The best we can do is single out a variable by the index of its binding site (the
DeBruijn index in this case), but with \texttt{METIS} potentially generating hundreds of
fresh variables during resolution, it would be disastrous to take such indices as a
direct representation of \texttt{METIS} variables.

Instead, we introduce maps from bound variable indices to \texttt{METIS} variables and carry
the mapping around in our clauses. A clause of \(n\) variables and \(m\) literals is
therefore represented by a pair consisting of a list of \texttt{METIS} variables and a
ProofScript theorem:
$$([v_1, v_2, v_3, \ldots, v_n], \vdash\forall x_1\;x_2\;\ldots\;x_n.\; \phi_1 \vee \phi_2
\vee \ldots \vee \phi_m).$$

where we assume that the \(i^{\text{th}}\) bound variable corresponds to the \texttt{METIS}
variable \(v_i\). Plain integers make a convenient choice for the representation of
these \texttt{METIS} variables, and at the call-site, we ask the resolution prover to use
them both internally and in proof certificates. For a concrete example, the \texttt{METIS}
clause \(\{p(3,8), \neg q(2,5)\}\) could be represented by any of the ProofScript pairs:
\begin{align*}
  &([3, 2, 8, 5], \vdash\forall x\;y\;z\;w.\; p\ x\ z \vee \neg q\ y\ w);\\
  &([8, 5, 3, 2], \vdash\forall a\;b\;c\;d.\; \neg q\ d\ b \vee (p\ c\ a));\\
  &([5, 2, 8, 3], \vdash\forall x_1\;x_2\;x_3\;x_4.\; p\ x_4\ x_3 \vee \neg q\ x_2\ x_1).
\end{align*}
Now suppose we want to perform the substitution \([h(1,2,9)/8]\). The resulting clause
will be \(\{p(3,h(1,2,9)), \neg q(2,5)\}\), represented by a ProofScript clause that is
$\alpha$-equivalent to

$$\vdash\forall x\;y\;z\;u\;v.\; p\ x\ (h\ y\ z\ u) \vee \neg q\ z\ v.$$

To find this, we must first figure out how many bound variables will be needed after the
substitution, generate fresh constants for them, and then recreate the mapping from
indices to \texttt{METIS} free variables.

Next suppose we want to resolve our new clause with \(\{q(2, 5), r(4, 6)\}\) which might
happen to be represented by

$$([5,4,2,6], \vdash\forall x\;y\;z\;w.\; r\ y\ w \vee q\ z\ x).$$

We can strip away quantifiers from one of the clauses to get at the literals,
introducing new constants into the current context along the way, but we will have to
make sure that we instantiate rather introduce variables in the other clause whenever
our bound variable mappings tell us there is a match. And after resolution, we will
have lost the free \texttt{METIS} variable \(5\), and so will be deleting it from our index
mapping. 

This requires a lot of delicate work, and a clear understanding of how the
introduction of universal quantifiers by lifting out of contexts can be exploited
programmatically. It is not the sort of thing we expect the average ProofPeer user to
get their hands dirty with, though if they feel the need to, the tight fit between
ProofScript's structured proof language and its programming language should make it
seem natural. But the regular user, we hope, will mostly enjoy writing simple
theorems such as the following, based on our working implementation of \texttt{METIS}:

\lstset{label= ,caption= ,captionpos=b,numbers=none,mathescape=true}
\begin{lstlisting}
let oneDef:'one = $\mathcal{P}$ $\emptyset$'
let twoDef:'two = $\mathcal{P}$ one'

theorem one:'$\forall x. x \in \texttt{one} = (x = \emptyset)$'
  by metis [empty,oneDef,power,subset,ext]

theorem two:'$\forall x.\; x \in \texttt{two} = (x = \emptyset \vee x = \texttt{one})$'
  by metis [empty,one,twoDef,power,subset,ext]

theorem oneNotZero: '$\neg(\emptyset = one)$'
  by metis [empty, one]
\end{lstlisting}

\section{Conclusion}
\label{sec:orgf5b06aa}
The scripting language for \emph{ProofPeer}, aptly named \emph{ProofScript}, is intended to be
the programmable metalanguage for an LCF kernel, but what distinguishes it from other
languages designed for this purpose, such as Standard ML, is that it aims primarily
to be the natural vehicle for writing accessible, \emph{declarative} proofs, where
black-box automation is left to fill out the tedious gaps.

The route we have taken towards this goal has not required any stops at the tactic systems
or contextual rewriters of traditional LCF, but we have made extensive use
of conversions and higher-order conversionals in order to connect to the \texttt{METIS}
first-order theorem prover. If the success of the Isabelle \emph{Sledgehammer} project is
anything to go by, this could be the only integrated tool that we need when it comes
to first-order automated proof.

ProofScript shares qualities of the ML dialects in its succinctness and support for
higher-order functions, and we were pleased that we were able to reason easily with
what would become quite complex types in ML. We did not really miss the static type
system reminding us not to mix list parameterised conversionals with term
parameterised ones. This is not to say that there are not other benefits to having a
type system, in terms of automatic documentation in type-signatures and in supporting
refactoring of complex code-bases, but our needs are quite modest for now.

When designing the logic of ProofScript, we aimed at some conceptual simplifications
so as not to distract users. Users are not left to choose between building
mathematical structures out of user-defined types and exhibiting those structures in
ZFC sets as they are in HOL/ZFC~\cite{HOLZFC}. Our type inference~\cite{TypeInferenceZFH} algorithm guides users away from
polymorphic universes to the ZFC one. And users writing declarative proofs are not
left wondering what the difference is between a universally quantified theorem and
one where all the variables are free to be instantiated. The distinction between free
variables and constants is just a logical confusion to a user working in a
declarative setting where constants can be freely introduced in local contexts.

But when it comes to the nuts-and-bolts of \emph{programming} proofs, there is sometimes
another story. Simple polymorphic types can improve the efficiency of proofs by
replacing infinitely many metatheorems with a single instantiable one. And free
variables, effectively being named holes in terms and theorems which can be
identified across contexts, become a means by which a programmer can represent their
own instantiation mechanisms. Not having such things meant that in
sections~\ref{sec:org5a040c6}
and~\ref{sec:org81faae6}, we were forced to shadow DeBruijn
indices with our own metadata. This is just the usual reminder that there are
balances to be struck between conceptual simplicity and practicalities. We feel
justified that we have struck such a balance.

Future work on automation in ProofPeer will focus on replicating the impressive and
ongoing development of Sledgehammer and its HOL~Light counterpart HOL(y)
Hammer~\cite{HOLyHammer}, thus allowing Proofscript to access a wide variety of powerful ATP tools
which can then be used to advise the \texttt{METIS} prover. As we proceed, we expect that
the necessary infrastructure of ProofScript will remain lean compared to those of its
friends in the LCF tradition.

\bibliographystyle{plain}
\bibliography{proofpeer}{}
\end{document}